\documentclass[1p,times,procedia]{elsarticle}
\usepackage{SubjetsQM2015}

\volume{00}
\firstpage{1}
\journalname{Nuclear Physics A}
\runauth{X.~Zhang, L.~Apolin\'ario, J.~G.~Milhano and M.~P{\l}osko\'n}

\jid{nupha}
\jnltitlelogo{Nuclear Physics A}

\begin{document}

\begin{frontmatter}

\dochead{}

\title{Sub-jet structure as a discriminating quenching probe}

\author[lbnl]{X.~Zhang}
\author[cent]{L.~Apolin\'ario}
\author[cent,cern]{J.~G.~Milhano}
\author[lbnl]{M.~P{\l}osko\'n}

\address[lbnl]{Lawrence Berkeley National Laboratory, 1 Cyclotron Road, Berkeley, CA 94720, USA}
\address[cent]{CENTRA, Instituto Superior T\'ecnico, Universidade de Lisboa, Av. Rovisco Pais, P-1049-001 Lisboa, Portugal}
\address[cern]{Physics Department, Theory Unit, CERN, CH-1211 Gen\`eve 23, Switzerland}

\begin{abstract}
In this work, we propose a new class of jet substructure observables which, unlike fragmentation functions, are largely insensitive to the poorly known physics of hadronization.
We show that sub-jet structures provide us with a large discriminating power between different jet quenching Monte Carlo implementations.
\end{abstract}


\end{frontmatter}


\section{Motivation}\label{sec:Intro}

Jets are one of the most important tools to characterize the properties of the hot and dense deconfined matter, the quark gluon plasma (QGP), formed in ultra-relativistic heavy-ion collisions, via the medium modifications of jet fragmentation relative to pp collisions (jet quenching).
The observations made at RHIC and LHC via the measurement of inclusive jet production at high transverse momentum ($\pT$), dijet energy imbalance and hadron/$\gamma$--jet coincidences have shown that jets are strongly modified in a QCD medium~\cite{Akiba:2015jwa}.
To further understand the details of such interactions, a well defined observable, strongly sensitive to internal jet structure modifications, is required.
In this proceeding, the medium modified internal jet structure is explored through the use of sub-jets which are obtained by re-clustering the constituents of a given jet.
The sub-jet production is sensitive to the details of jet fragmentation and mostly immune to the uncertainties arising from the poorly understood physics of hadronization.
Since the catchment area of sub-jet is smaller than the original jet, background fluctuations are reduced in sub-jets relative to original jets.
\section{Setup}\label{sec:Setup}

\begin{figure}[t]
\begin{center}
\includegraphics[width=0.43\textwidth]{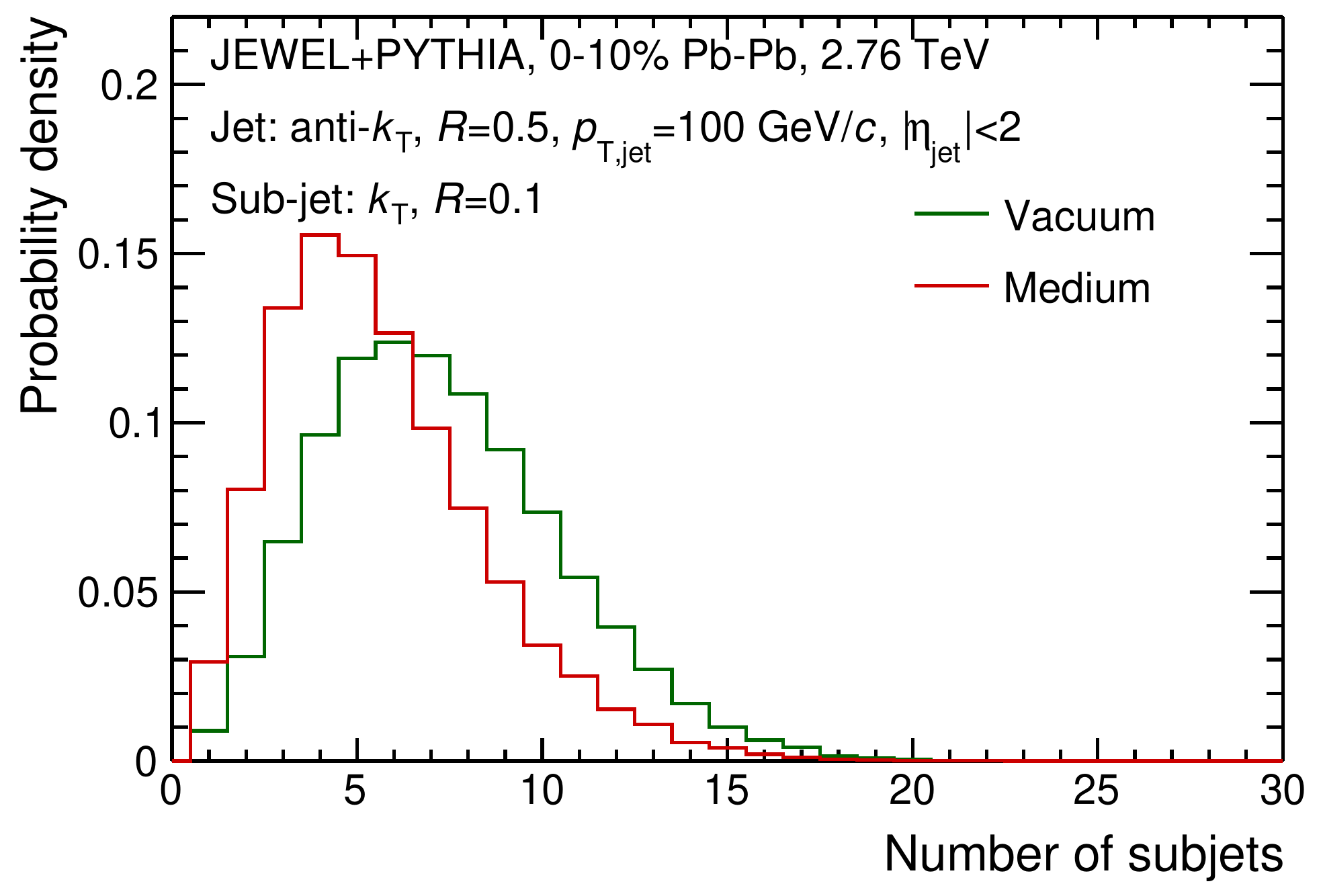}
\includegraphics[width=0.43\textwidth]{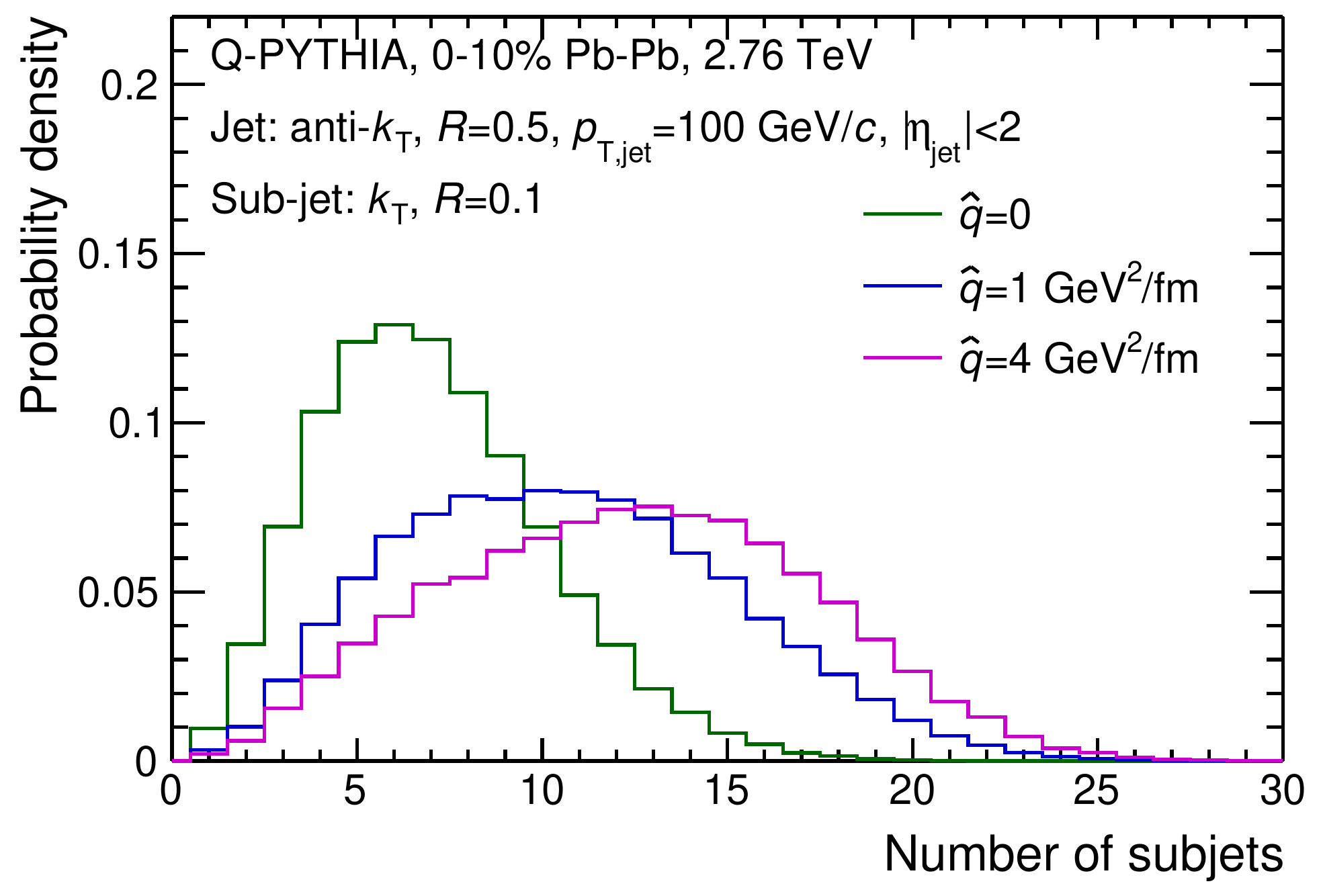}
\end{center}
\vspace{-0.5cm}
\caption{Sub-jet multiplicity distribution in jets with $\pT=100~\GeVc$ from \textsc{Jewel} (left) and \textsc{Q-Pythia} (right). The results are compared to corresponding vacuum references, see the text for  details.}
\label{fig:Nsj}
\end{figure}

Two different quenching models are used in this investigation.
\begin{itemize}
\item \textsc{Q-Pythia}~\cite{Armesto:2009fj}: the medium effects are introduced by considering BDMPS-Z type radiation energy loss via the modification of the vacuum splitting function.
It results in soft and large angular gluon radiation.
The quenching strength is described by the transport coefficient, $\qhat$, defined as the average transverse momentum squared acquired per mean free path length of a parton traveling in the medium.
\item \textsc{Jewel}~\cite{Zapp:2012ak}: contains both radiative and elastic energy loss.
The medium induced radiation is generated by the parton shower,
while parton in-medium re-scattering is described by leading order $2\to 2$ matrix elements.
The Landau-Pomeranchuck-Migdal effect is also included in the description of gluon emission.
\end{itemize}
For each model, the events are simulated in the conditions of Pb--Pb collisions at $\sNN=2.76$~TeV in the \cent{0}{10} centrality class.
The vacuum reference of \textsc{Jewel} is provided by its vacuum mode.
For \textsc{Q-Pythia}, the vacuum reference is obtained with $\qhat=0$.

The FastJet package~\cite{Cacciari:2011ma} is used to reconstruct the jets and sub-jets.
The jets are reconstructed by using all final state stable particles in the pseudo-rapidity window $\abs{\eta}<2.6$ with anti-$\kT$ algorithm~\cite{Cacciari:2008gp} and resolution parameter $R=0.5$.
To avoid fiducial effect, the reconstructed jets are selected in $\abs{\eta}<2$.
A list of sub-jets is obtained by reclustering the constituents in a given jet with a smaller $R$ and a different jet reconstruction algorithm.
In this analysis, the  sub-jets are obtained with $R=0.1$ and $\kT$ algorithm~\cite{Ellis:1993tq}.
\section{Sub-jet structure}\label{sec:c02Subjets}

\subsection{Sub-jet multiplicity}

The distributions of the number of sub-jets in jets with $\pT=100~\GeVc$ from \textsc{Jewel} and \textsc{Q-Pythia} with $\qhat=1$~GeV/fm$^{2}$ and $4$~GeV/fm$^{2}$, respectively, are shown in the left and right panels of figure~\ref{fig:Nsj}.
In each case, the results in medium are compared to the corresponding vacuum reference.
In \textsc{Jewel}, the mean multiplicity of sub-jets in medium is smaller than in the vacuum,
while an opposite trend is observed with \textsc{Q-Pythia}: it gives on average larger sub-jet multiplicity in the medium than in the vacuum, the distribution in medium is much broader than the vacuum reference and it evolves with quenching strength.
The results show that the sub-jet multiplicity is quite sensitive to quenching scenarios as well as the quenching strength.
It is worth to mention that these results are obtained without considering the combinatorial background in heavy-ion collisions which will contaminate the sub-jet production.
The identified difference of sub-jet multiplicity distributions between the quenching models will be weakened by the presence of heavy-ion background.

\subsection{Difference between the two hardest sub-jet $\pT$}

\begin{figure}[t]
\begin{center}
\includegraphics[width=0.43\textwidth]{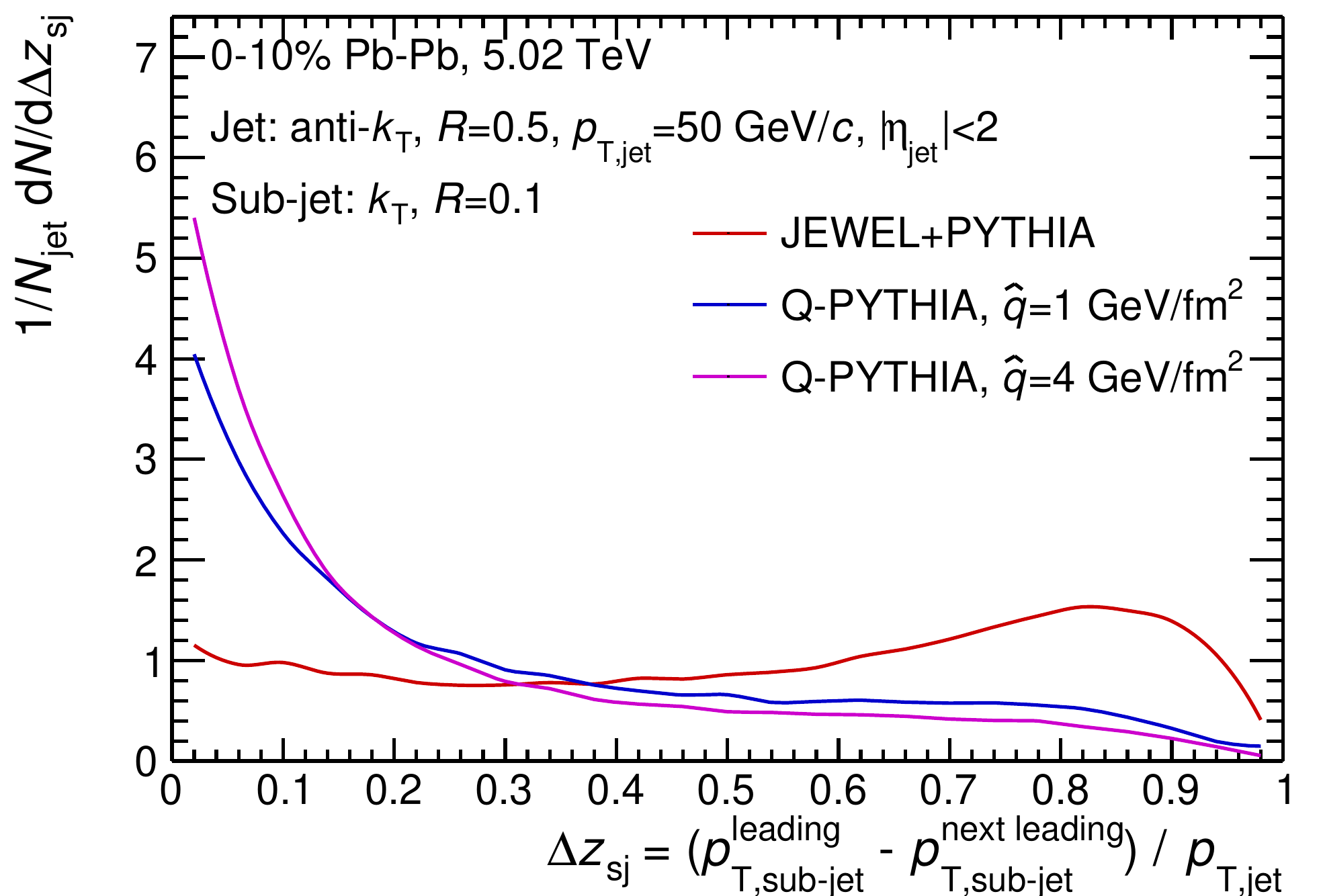}
\includegraphics[width=0.43\textwidth]{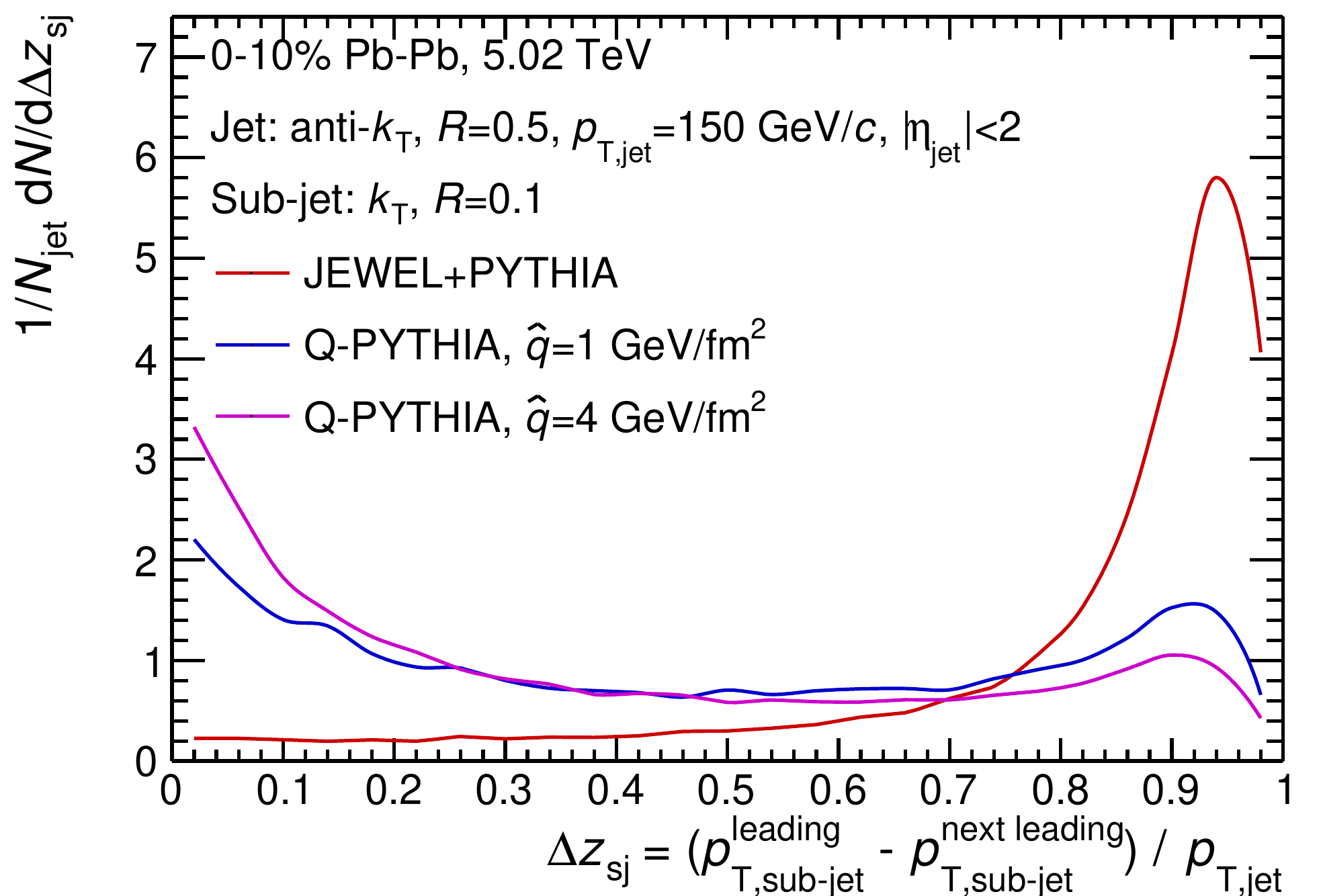}
\end{center}
\vspace{-0.5cm}
\caption{$\Zsj$ distributions from \textsc{Jewel} and \textsc{Q-Pythia} in jets with $\pT=50~\GeVc$ (left) and $150~\GeVc$ (right), see the text for details.}
\label{fig:Dzsj}
\end{figure}

To correct for the large combinatorial background in heavy-ion collisions, a generic applied approach was proposed in~\cite{Cacciari:2007fd}:
\begin{equation}\label{eq:bkgsub}
\pT[sub-jet]=\pT[sub-jet]^{\rm raw} - \rho A\pm\delta_{\rm bkg},
\end{equation}
where $\pT[sub-jet]^{\rm raw}$ and $\pT[sub-jet]$ refer to the measured and corrected sub-jet $\pT$, respectively, $A$ is the sub-jet area, $\rho$ is the event-by-event background density, and $\delta_{\rm bkg}$ is the deviation between $\pT^{\rm corr}$ and true $\pT$ due to local $\rho$ fluctuations.
By considering the two hardest sub-jets in a given jet, which are close to each other and thus suffer very similar background fluctuations, the term $\delta_{\rm bkg}$ in eq.~(\ref{eq:bkgsub}) will be largely cancelled in the difference between their $\pT$:
\begin{equation}
\Delta\pT[sj] = \pT[sub-jet]^{\rm leading} - \pT[sub-jet]^{\rm next~leading},
\end{equation}
where $\pT[sub-jet]^{\rm leading}$ and $\pT[sub-jet]^{\rm next~leading}$ denote the $\pT$ of leading and next leading sub-jets, respectively.

Figure~\ref{fig:Dzsj} shows the distributions of $\Zsj$, defined as $\Zsj=\Delta\pT[sj]/\pT[jet]$,
in \textsc{Jewel} and \textsc{Q-Pythia} in jets with $\pT=50~\GeVc$ and $150~\GeVc$ in the left and right panels, respectively.
\textsc{Jewel} gives a large energy imbalance between the two hardest sub-jets, and a peak is observed at $\Zsj\to 1$, this effect being more significant with increasing $\pT[jet]$.
In contrast, the two hardest sub-jets $\pT$ are more balanced in \textsc{Q-Pythia}, and this results in an enhancement at $\Zsj\to 0$.
Similarly as the sub-jet multiplicity distribution shown in figure~\ref{fig:Nsj}, the two models are well separated by the $\Zsj$ distribution in a very large $\pT[jet]$ range.

\subsection{Sub-jet tagged jets}

\begin{figure}[t]
\begin{center}
\includegraphics[width=0.43\textwidth]{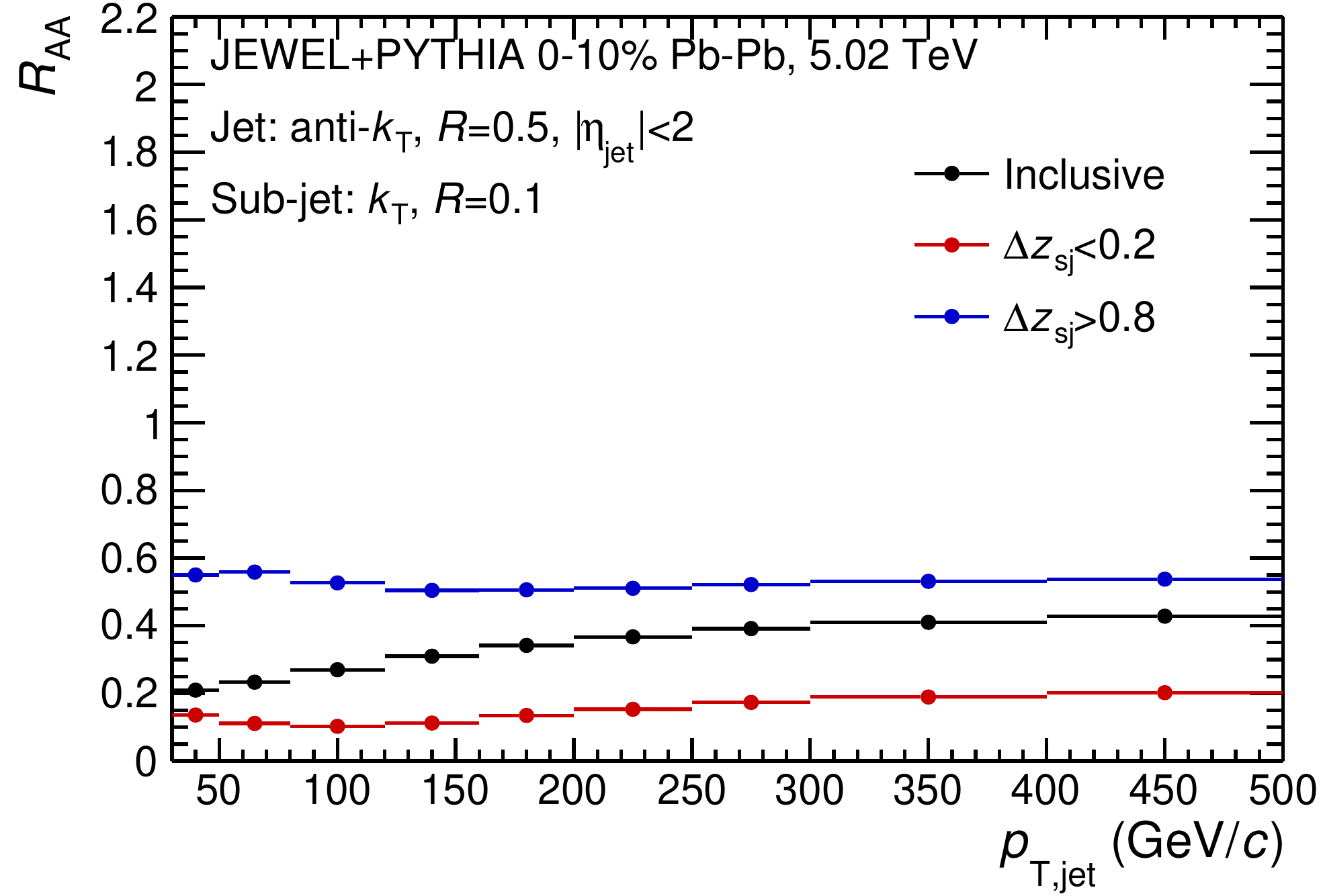}
\includegraphics[width=0.43\textwidth]{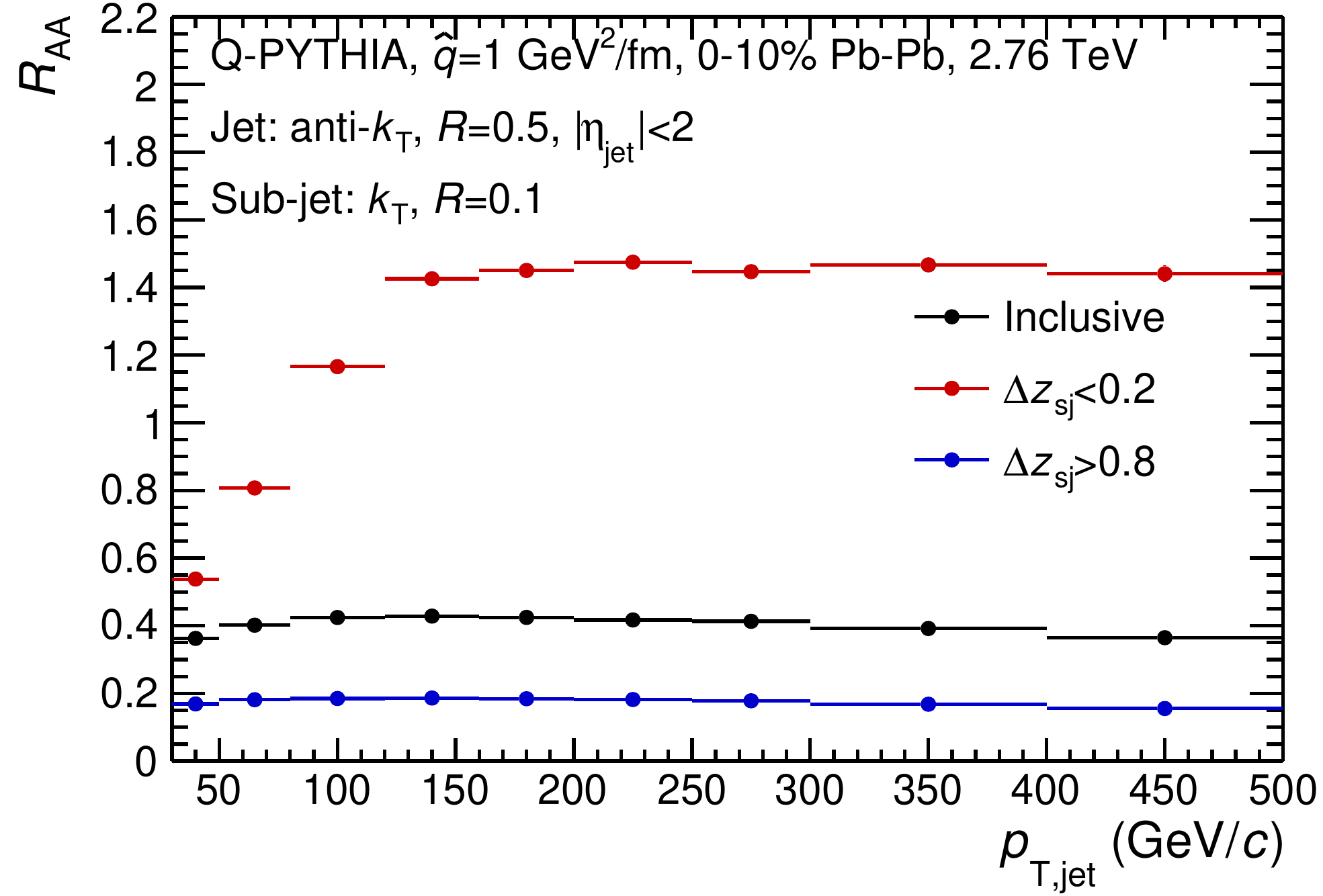}
\end{center}
\vspace{-0.5cm}
\caption{$\RAA$ of jets tagged by $\Zsj$ in $\Zsj<0.2$ and $\Zsj<0.8$ in \textsc{Jewel} (left) and \textsc{Q-Pythia} (right). The results are compared to the inclusive jets.}
\label{fig:RAA}
\end{figure}

The medium-modified jet production is quantified by the nuclear modification factor ($\RAA$), defined as the ratio of the jet yield measured in heavy-ion collisions to that observed in pp collisions scaled by the number of binary nucleon--nucleon collisions.
Since the $\Zsj$ observable is robust against heavy-ion background, it also provides an unbiased selection of the jet samples in heavy-ion and pp collisions.
The $\RAA$ of jets selected with $\Zsj<0.2$ and $\Zsj>0.8$ are shown in figure~\ref{fig:RAA}.
Compared to the inclusive jet $\RAA$, \textsc{Jewel} and \textsc{Q-Pythia} have opposite response to the $\Zsj$ selection.
The $\RAA$ of jets tagged by $\Zsj>0.8$ ($\Zsj<0.2$) is enhanced (suppressed) in \textsc{Jewel} and suppressed (enhanced) in \textsc{Q-Pythia}.
Especially in \textsc{Q-Pythia}, the $\RAA$ of jets selected in $\Zsj<0.2$ is $\sim 4$ times higher than the inclusive ones.
The measurement of the $\RAA$ of jets selected by $\Zsj$ in data will provide a clean way to differentiate the quenching scenarios.
This approach can also be applied for other jet measurements such as the per-trigger jet yield and dijet energy asymmetry.
\section{Conclusion}\label{sec:c04Concl}

The sub-jet structure, which is sensitive to the  medium-modified internal jet structure, has been explored as a tool to differentiate the quenching scenarios.
A new observable $\Delta\pT[sj]$ (or $\Zsj$) was proposed since it is sensitive to the medium modifications of the jet fragmentation and robust against the background fluctuations in heavy-ion collisions.
It also provides an unbiased selection of jet samples in heavy-ion and pp collisions, simultaneously.
It has been shown that, by measuring observables of jets tagged by $\Delta\pT[sj]$ (or $\Zsj$) will provide a clean way to reveal the jet quenching mechanism observed in data.

\section*{Acknowledgements}
This work of LA and JGM was partly supported by Funda\c c\~ao para a Ci\^encia e a Tecnologia (Portugal) under project SFRH/BPD/103196/2014.


\bibliographystyle{elsarticle-num}
\bibliography{SubjetsQM2015}

\end{document}